\documentstyle[epsfig]{mn}
\begin{document}


\title
[Distance measurements as a probe of cosmic acceleration] 
{Distance measurements as a probe of cosmic acceleration} 
\author[Neil Trentham] 
{
Neil Trentham$^1$\\
$^1$ Institute of Astronomy, Madingley Road, Cambridge, CB3 0HA.\\
}

\maketitle 
\begin{abstract} 
A major recent
development in observational cosmology has been an accurate
measurement of the 
luminosity distance-redshift relation out to redshifts $z=0.8$
from Type Ia supernova
standard candles.  The results have been argued as evidence for
cosmic acceleration.  It is well known that
this assertion depends on the assumption that we
know the equation of state for all mass-energy other than normal pressureless
matter; popular models are based on either the cosmological constant or on
the more general quintessence formulation.
But this assertion also depends on a number of other assumptions, implicit in
the derivation of the standard cosmological field equations: large-scale
isotropy and homogeneity, the flatness of the Universe, and the validity
of general relativity on cosmological scales (where it has not been
tested).  A detailed
examination of the effects of these assumptions on the interplay between
the 
luminosity distance-redshift relation and the acceleration of the Universe is
not possible unless one can define the precise nature of the failure of
any particular assumption.  However a simple quantitative investigation is
possible and reveals a number of considerations about the relative importance
of the different assumptions.  In this paper we present such an
investigation.   
We find that the relationship between the distant-redshift
relation and the sign of the deceleration parameter is fairly
robust and is 
unaffected if only one of the assumptions that we investigate is invalid
so long as the deceleration parameter is not close to zero (it would
not be close to zero in the currently-favored  $\Omega_{\Lambda} =
1 - \Omega_{\rm matter}$ = 0.7 or 0.8 Universe, for example).  
Failures of two or more assumptions in concordance
may have stronger effects.
\end{abstract} 


\section{Introduction and Background}

One of the most significant cosmological results of recent times 
has been an accurate determination of the luminosity distance -- redshift 
relation (Perlmutter et al.~1999, Schmidt et al.~1998, Riess 1999,
Zehavi \& Dekel 1999, Riess et al.~2000, Burrows 2000), 
from measurements of Type Ia supernova standard candles 
out to redshifts  $z \sim 0.8$. 
The measurements are highly inconsistent with a flat purely matter-dominated
Universe and have consequently
been regarded as evidence for cosmic acceleration.

But this assertion depends intricately on other assumptions made, and does not
necessarily 
follow from the measurement of the luminosity distance-redshift relation.
Suppose that we can describe the expansion of the Universe by a scale-factor
$R (t)$ which is a function of cosmic time $t$.  The redshift $z
= R(t_0)/R(t) - 1$  is then also a function of $t$ (here $t_0$ is
the current age of the Universe).  
The luminosity distance-redshift 
relation 
is (e.g.~Weinberg 1972) 
\begin{equation}
d_L (z) = R(t_0) (1+z) S_k \left( \int_{R(t_0)\over(1+z)}^{R(t_0)}
{{c \,\,{\rm d}R} \over { {{\rm d}R\over{\rm d}t} \, R}} \right),
\end{equation}
where $c$ is the speed of
light, $k$ is an additional parameter describing the
curvature of the Universe ($k=+1$ for a closed Universe, $k=0$ for a
flat Universe, and $k=-1$ for an open Universe), and $S$ is defined so that
$S_{+1} (x) = \sin x$,
$S_0 (x) = x$, and $S_{-1} (x) = \sinh x$.   
The deceleration parameter is (e.g.~Weinberg 1972) 
\begin{equation}
$$q_0 = - 
\left.{
{{{{ {\rm d}^2 R}\over{{\rm d} t^2}} R  }\over {\left({{\rm d}R\over{\rm d}t} 
\right)^2}}}\right|_{t_0}.
\end{equation}
If the universe is accelerating $q_0 < 0$.
It follows from these equations that a function $R(t)$ constrained 
from a measurement of $d_L (z)$ does not uniquely define a value of $q_0$.
In fact, it does not even uniquely define the {\it sign} of $q_0$.
This happens because $q_0$ depends on the second time derivative of $R(t)$
but $d_L(z)$ only depends on lower order derivatives and their integral. 
As a simple (albeit unphysical)
example, consider a function $R(t)$ that can be
expressed as a polynomial in $t$ with coefficients $a_n$ such that
$R(t) = \Sigma a_n t^n$ where the $n$th term contributes significantly more 
than
the $(n+1)$th one.  If one then defines another function $R^{\prime} (t)$ 
that is identical to $R(t)$ except that the sign of the $a_2$ coefficient
is opposite, then both $R(t)$ and $R^{\prime} (t)$ generate almost identical 
$d_L (z)$ relations but deceleration parameters with opposite signs.

It is conventional to draw a connection 
between the $d_L (z)$ relation and the sign of the
deceleration parameter as follows.
One assumes the 
Friedmann-Robertson-Walker metric (which follows from the requirement
of isotropy and homogeneity on the largest scales), 
Einstein's theory of general relativity,
zero spatial curvature ($k=0$) 
and an equation of state so that the only contributions to the stress-energy
tensor in general relativity comes from normal matter and the vacuum energy
(the cosmological constant).  This latter assumption is frequently 
examined (Caldwell, Dave \&
Steinhardt 1998; Garnavich et al.~1998; Efstathiou 1999;
Wang et al.~2000; Maor, Brustein \& Steinhardt 2000) 
since it has no theoretical basis. 
If these assumptions are made, 
the model forces a particular form of $R(t)$ and consequently
a particular form of $d_L(z)$ and $q_0$ 
in terms of $\Omega_M$ and 
$\Omega_{\Lambda}$, the cosmological
densities of matter and the vacuum.
The deceleration parameter here has the particularly simple form
$q_0 = \Omega_M / 2 -
\Omega_{\Lambda}$.  
Given the constraints imposed by the supernova measurements,
it is found that the only pairs of $\Omega_M$ and $\Omega_{\Lambda}$ that
are allowed force $q_0 < 0$.  Therefore the 
Universe is argued to be accelerating.   

But any of the assumptions may turn out to be invalid.  Aside from the
much-examined assumptions regarding the equation of state turning out
to be untrue, it is also 
possible that $k \neq 0$, that very-large-scale 
inhomogeneities may exist (C{\'{e}}l{\'{e}}rier 2000),
or even that general relativity may turn out to be
invalid on the very largest scales (where is has not been directly tested). 
If any or all of these turn out to be true, 
we might no longer require an accelerating Universe given
the current (or even a far more precise)
measurement of the $d_L(z)$ relation; this possibility
is investigated in a simple
quantitative way in this paper. 

\section{Friedmann models}

The formulation that is normally used to define the cosmological parameters
can be conceptually (albeit arbitrarily) 
divided into the following four steps.
\vskip 1pt
\noindent
(i) The cosmic geometry is described by the four-dimensional Riemann
curvature tensor.  If isotropy and homogeneity are assumed, the resulting
constraints on the Riemann tensor force the space-time metric to have
a specific form, given by the Friedmann-Roberston-Walker metric.  
Implicit in this form is the parameter $k$, which takes the value 0 or
$\pm 1$.  The existence of
this parameter $k$ follows from the assumption that the Ricci scalar (the
one-dimensional fully-contracted Riemann tensor) is not a function of
either time or of position, which is required by 
homogeneity. 
If this assumption about homogeneity and isotropy is made, then observations
of the position $l_{\rm peak}$ 
of the first Doppler peak in the microwave background
spectrum (e.g.~de Bernardis et al.~2000) that suggest a value of the total
cosmological 
density $\Omega_{\rm TOT}$ close to 1 in units of the critical density
then suggest that $k=0$.  
\vskip 1pt
\noindent
(ii) From the metric, one can then compute explicitly the curvature
tensor and its contractions.  From the equations of general relativity,
one can then compute the cosmological field equations.  
Note that general relativity has not been verified observationally to
high precision on these
cosmological scales (although it is clearly very
successful on smaller scales -- see Section 8 of Weinberg 1972)
so its application here is very much an assumption.
\vskip 1pt
\noindent
(iii) if we then adopt an equation of state that relates density to
pressure for every contribution of mass density in the field equations, 
we can then derive $R(t)$ as a function of these contributions.  
Some common strategies are to assume that
(a) the only contribution comes from normal pressureless matter;
(b) part of the contribution comes from normal matter and all the remaining
part comes from a cosmological constant; (c) 
part of the contribution comes from normal matter and all the remaining 
part comes from some material whose equation of state is defined by the
quintessence formulation of Caldwell et al.~(1998).  
\vskip 1pt
\noindent
(iv) from measurements of 
luminosity distance-redshift relation we constrain 
combinations of parameters that arise from the considerations in (iii)
above.  For example, if one has specified exactly two forms of mass density,
both of which have known equations of state, then one can constrain the
relative mass densities of the two components (this would be true in case
(b) above).    
If one makes a further
assumption, like requiring a flat
Universe ($k=0$), then one has another constraint and in conjunction with
the luminosity
distance-redshift relation can provide a precise determination of the
model parameters. 
Alternatively any related measurement like the value of $l_{\rm peak}$  
will have a similar effect (recall from (i) that in the context of these 
models, the value of $l_{\rm peak}$ that was measured is operationally
similar to requiring $k=0$).  In practice, the constraints on the model
parameters in case (b) are tight since the constraints on the  
mass densities in normal matter and in the cosmological constant are
almost orthogonal (see Efstathiou et al.~1999).  
From the permitted values of the contributions of these mass densities,
we can then compute $R(t)$ and hence
$q_0$, within the errors provided by the
measurements.  

\begin{table*}
\caption{Simple modifications to the Friedmann models}
{\vskip 0.75mm}
{$$\vbox{
\halign {\hfil #\hfil && \quad \hfil #\hfil \cr
\noalign{\hrule \medskip}
Modification and form & New Friedmann equation$^{\rm a}$ & \cr 
  & & \cr
  & & \cr
\noalign{\smallskip \hrule \smallskip}
\cr
Isotropy and homogeneity & &\cr
 ${\cal R} \sim r^{\epsilon} $ & as standard &\cr 
   & & &\cr
  ${\cal R} \sim t^{\epsilon} $ & as standard &\cr 
   & & &\cr
    ${\cal R} \sim \sin^{\epsilon}\theta $ &
$  (-\epsilon - \epsilon\cos{2 \theta} + 2 \sin^2{\theta}) \,\, \epsilon
\,\, \sin^{-2+\epsilon}{\theta}\,\, + $ &\cr 
  & $4 r^2 (1 + 3 \sin^{\epsilon}{\theta} - \sin^{2 \epsilon}{\theta})
  {\dot{R}}^2
 = 32 \pi G \rho r^2 R^2 $ &\cr
   & & &\cr
             ${\cal R} \sim \sin^{\epsilon}\phi $ &
$  (2 - \epsilon - \epsilon \cos{2 \phi}) \,\, \epsilon
\,\, {\rm cosec}^2{\theta} \,\, \sin^{-2+\epsilon}{\phi}\,\, + $ &\cr 
 &  $4 r^2 (1 + 3 \sin^{\epsilon}{\phi} - \sin^{2 \epsilon}{\phi}) {\dot{R}}^2
 = 32 \pi G \rho r^2 R^2 $  &\cr
  &  &\cr
  &  &\cr
Density (curvature)  & &\cr
 $\Omega_M + \Omega_{\Lambda} = 1 + \epsilon$ &
as standard & \cr
  &   &\cr
  &   &\cr
General relativity & &\cr
$R_{\mu \nu} - {1\over{2}} g_{\mu \nu} {\cal{R}} =
8 \pi G T_{\mu \nu} (T^{\alpha}_
{\,\,\alpha})^{\epsilon}$
& $H^2 = H_0^2 \left({R\over{R_0}}\right)^{-3 (1+\epsilon)}$ &\cr 
  &   &\cr
  &   &\cr
Equation of state &  &\cr
$P = \epsilon \rho$ (quintessence$^{\rm b}$) & $H^2 = \Omega_M
H_0^2 \left({R\over{R_0}}\right)^{-3} +
  (1-\Omega_M) H_0^2 \left({R\over{R_0}}\right)^{
-3 (1 + \epsilon)}$ &\cr
\noalign{\smallskip \hrule}
\noalign{\smallskip}\cr}}$$}
\begin{list}{}{}
\item[$^{\mathrm{ }}$] The symbols have the following meanings:
$\epsilon$ is a parameter, defined individually in each case, that
describes how we are modifying the model; $(t,r,\theta,\phi)$ are the
time, radial, and two angular coordinates used to define the space-time
metric $g_{\mu \nu}$;  the two-dimensional contraction of the full
four-dimensional Riemann curvature tensor is the Ricci tensor
$R_{\mu \nu}$ and its contraction to one dimension is the Ricci scalar
${\cal R}$;
$T_{\mu \nu}$ is the Einstein stress-energy tensor, implicit in which
are the density $\rho$ and pressure $P$;
$R$ is the cosmic scale factor, which depends on $t$, ${\dot{R}}$ is its
time derivative and $H = {{\dot{R}}\over{R}}$;
$R_0$ and $H_0$ (the Hubble constant) are the
present-day values of $R$ and $H$;
$\Omega_M$ is the cosmological density of normal matter obeying the
equation of state $P=0$ in units of the critical density;
$\Omega_{\Lambda}= \Lambda/3 H_0^2$ is the cosmological density in the
cosmological constant;  
$G$ is the gravitational constant.
Other symbols are defined elsewhere in the text;
the speed of light $c$ is set to unity in all the equations in the table.
\item[$^{\mathrm{ }}$]
\item[$^{\mathrm{a}}$] Assuming $k=0$ and $\Lambda=0$.
\item[$^{\mathrm{b}}$] In order to fully define a quintessence model we need
to state both the equation of state of the 
quintessent material (parameterized
by $\epsilon$ here) and state the density of this material.  
For the
models we consider here, we fix the density of normal matter to be 
$\Omega_M =0.2$ in
units of the critical density and the density of the 
quintessent material
to be $\Omega_{\epsilon} =0.8$
so that the total density is 
$\Omega_{\rm TOT} =1$. 
The value of $\Omega_M =0.2$ 
is suggested by dividing the value of $\Omega_{\rm baryon}$
derived from big bang nucleosynthesis constraints (e.g.~Smith, Kawano \&
Malaney 1993) by the baryon fraction in rich galaxy clusters like Coma
(White et al.~1993).  It is also the value suggested for $\Omega_{\rm dark
\,\,matter}$ by (see Trimble 1987)
assigning an amount of dark matter to each luminous galaxy
(e.g.~using the correlations of Kormendy 1990) and integrating over the
galaxy luminosity function (e.g.~Ellis et al.~1996).
Neither of these methods depend on the presence or absence of a form of
mass density that does not behave according to the equation of state $P=0$,
like a cosmological constant.
The value of $\Omega_{\rm TOT} =1$ is suggested by the measurements
of the cosmic microwave background by de Bernardis et al.~(2000).
 
\end{list}
\end{table*}

\begin{table*}
\caption{Fiducial models}
{\vskip 0.75mm}
{$$\vbox{
\halign {\hfil #\hfil && \quad \hfil #\hfil \cr
\noalign{\hrule \medskip}
Model  & Parameters & $k$  & $d_L (z=1) / c H_0^{-1}$ & $ q_0 $ &\cr
 &  & & & & \cr
\noalign{\smallskip \hrule \smallskip}
\cr
A & $\Omega_M = 1$  & 0 & 1.17 & 0.50 & \cr
B & $\Omega_M = 0.2$  & $-$1  & 1.41 & 0.10 & \cr
C & $\Omega_M = 0.2$, $\Omega_{\Lambda} = 0.8$  & 0 & 1.65 & $-0.70$ & \cr
\noalign{\smallskip \hrule}
\noalign{\smallskip}\cr}}$$}
\end{table*}

\begin{table*}
\caption{Effects of modifications on observables}
{\vskip 0.75mm}
{$$\vbox{
\halign {\hfil #\hfil && \quad \hfil #\hfil \cr
\noalign{\hrule \medskip}
Modification and form & 
 Values of $\epsilon$ that  & Values of $\epsilon$ that  & 
 Values of $\epsilon$ that   & \cr 
  & are consistent with a & are consistent with the  & keep unchanged &\cr
  & measurement of $d_L (z=1)$  & fiducial value of $q_0$    
&    the sign of $q_0$       &\cr
  &  that is accurate to 10\%    & to within 10\% &           &\cr
\noalign{\smallskip \hrule \smallskip}
\cr
Isotropy and homogeneity & & & &\cr
 ${\cal R} \sim r^{\epsilon} $ & any $\epsilon$ &
   any $\epsilon$ & any $\epsilon$ &\cr
   & & &\cr
             ${\cal R} \sim t^{\epsilon} $ & any $\epsilon$ &
     any $\epsilon$ & any $\epsilon$ &\cr
   & & &\cr
             ${\cal R} \sim \sin^{\epsilon}\theta $ &
depends on $(r,\theta)$ &\ 
depends on $(r,\theta)$ & depends on $(r,\theta)$ &\cr
   & & &\cr
             ${\cal R} \sim \sin^{\epsilon}\phi $ &
 depends on $(r,\theta,\phi)$ &
depends on $(r,\theta,\phi)$ & depends on $(r,\theta,\phi)$ &\cr
  & & & &\cr
  & & & &\cr
Density (curvature) & & & &\cr
$\Omega_M + \Omega_{\Lambda} = 1 + \epsilon$  
& A: $-0.45 < \epsilon < 0.63$
& $-0.10 < \epsilon < 0.10$
& $\epsilon > -1.00$ &\cr 
& B: $-0.30 < \epsilon < 0.41$
& $-0.02 < \epsilon < 0.02$
& $\epsilon > -0.20$ &\cr
& C1$^{\rm a}$: $-0.18 < \epsilon < 0.54$
& $-0.14 < \epsilon < 0.14$
& $\epsilon < 1.40 $ &\cr
& C2$^{\rm a}$: $-0.50 < \epsilon < 0.18$
& $-0.07 < \epsilon < 0.07$
& $\epsilon > -0.70$ &\cr
  & & & &\cr
  & & & &\cr
General relativity & & & &\cr
$R_{\mu \nu} - {1\over{2}} g_{\mu \nu} {\cal{R}} =
8 \pi G T_{\mu \nu} (T^{\alpha}_
{\,\,\alpha})^{\epsilon}$
& A: $-0.19 < \epsilon < 0.22$
& $-0.03 < \epsilon < 0.03$
& $\epsilon > -0.33$ &\cr
& B: $-0.89 < \epsilon < 0.52$
& $-0.03 < \epsilon < 0.03$
& $\epsilon > -0.33$ &\cr
& C: $-0.39 < \epsilon < 0.36$
& $-0.23 < \epsilon < 0.23$
& $\epsilon < 2.33 $ &\cr
  & & & &\cr
  & & & &\cr
Equation of state & & & &\cr
$P = \epsilon \rho^{\rm b}$
& A: $-0.295< \epsilon < 0.27$
& $-0.04 < \epsilon < 0.04$
& $\epsilon > -0.42$ &\cr
& B: $-0.79 < \epsilon < -0.21$
& $-0.34 < \epsilon < -0.33$
& $\epsilon > -0.42$ &\cr
& C: $-1.41 < \epsilon < -0.65$
& $-1.06 < \epsilon < -0.94$
& $\epsilon < -0.42$ &\cr
  & & & &\cr
\noalign{\smallskip \hrule}
\noalign{\smallskip}\cr}}$$}
\begin{list}{}{}
\item[$^{\mathrm{a}}$]For the fiducial model C, modifications that are
introduced can be considered as modifications to the behaviour of the
matter component (line C1) or the cosmological constant component (line C2).
Results are presented for both. 
\item[$^{\mathrm{b}}$]Note that the value of $\epsilon = 0$ is not
``permitted'' for models B and C.  This is because in neither case
can the fiducial model be described as an unmodified version of the
model under consideration here.  In fact, fiducial
model C is exactly
equivalent to a modified model with $\epsilon = -1$; this value of
$\epsilon$ is of course at the centre of the permitted range. 
The reason that models A, B, and C give different constraints on $\epsilon$
in the second and third columns is because
differrent fiducial models have been adopted.  In the fourth column, where
the limiting $\epsilon$ does not depend on the fiducial model, this limiting
value is the same in all three cases.  
\item[$^{\mathrm{ }}$]
\end{list}
\end{table*}

\section{Modifications to Friedmann models}

Clearly there are lots of assumptions involved.  
It is therefore useful to examine
each in detail and to see how relaxing each affects the interplay between
$d_L(z)$ and $q_0$.  
Assessing the general case is mathematically complicated and unconstrained.  
However, it is possible to compute the consequences of relaxing the
assumptions in simple well-defined ways, and this is done in Table 1.
The approach is as follows:
\vskip 1pt \noindent
(1) One of the assumptions described in the previous section is relaxed
(given in the first column of Table 1) and a modification of the relevant
equation is assumed.  Each adjustment is
described in terms of a parameter $\epsilon$, which is a measure of how
big the modification is.
\vskip 1pt \noindent
(2) The relevant modified Friedmann equation is derived (the second column of
Table 1).  
\vskip 1pt \noindent
(3) From this equation, we then compute the value of $d_L(z=1)$ for 
various fiducial
models, listed in Table 2 (note that Model C is consistent with the
observations described in Section 1, but Models A and B are not).  
These represent the case $\epsilon = 0$. 
We then ask: what values of $\epsilon$ generate
value of $d_L (z=1)$ that differs by less than ten per cent from this
fiducial value?   
These values are presented in Table 3. 
A measurement of the luminosity distance at $z=1$ that is precise to 10\%
requires observations slightly deeper and more precise than those of
Perlmutter et al.~1999, but should be attainable with larger samples of
supernovae observed with large telescopes in the near future.
\vskip 1pt \noindent
(4) We also compute the values of $\epsilon$ that leave the sign of the
deceleration parameter within 10 \% of its fiducial value and that
leave the sign of the
deceleration parameter unchanged 
(the third and fourth columns of Table 3). 
\vskip 1pt \noindent
If the values of $\epsilon$ allowed in the second column of Table 3 all
satisfy the condition given in the fourth column of Table 1, then we
can conclude that a measurement of $d_L(z=1)$ precise to 10\% does indeed
constrain the sign of $q_0$ (for example, this is true of the case
labeled ``General Relativity'') for the relevant fiducial model. 

An example of this calculation follows.  Consider the last row in
the ``density (curvature)'' section of the table i.e.~the line labelled ``C2''.
Given the Friedmann equation, 
equation (1) now
becomes 
\begin{equation}
d_L (z) = { {c (1+z)}\over{\sqrt{\vert \epsilon \vert } H_0}} 
\,\, S_k \left[  \sqrt{\vert \epsilon \vert} \, x_z (\epsilon) \right] 
\end{equation}
where
$k = +1$ if $\epsilon > 0$, $k = -1$ if $\epsilon < 0$ and 
$k = 0$ if $\epsilon = 0$, 
and
\begin{equation}
x_z (\epsilon) =   \int_{0}^{z} 
 { {{\rm d}z^{\prime}}\over{\sqrt{
 (1 + z^{\prime})^2 (1 + 0.2 z^{\prime}) + 
 (0.8 + \epsilon)  z^{\prime} (2 +  z^{\prime})}} }.
\end{equation}
If $\epsilon = 0$ (fiducial model C), then
$d_L (z=1) = 1.65 \,c / H_0$.  For a value within 10 per cent of this
($  1.48 \, c / H_0 < d_L (z=1) <  1.81 \, c / H_0$), Equation (3) requires
$-0.50 < \epsilon < 0.18$.   
Equation (2) now becomes
\begin{equation}
q_0 = -0.7 - 
\epsilon.
\end{equation}
For the fiducial model $q_0=-0.7$ and is negative.  
Values of $\epsilon$ that result in a value of $q_0$ within 10\% of this
value are $-0.07 < \epsilon < 0.07$, from Equation (5).  
To keep $q_0$ 
negative, Equation (5) requires $\epsilon > -0.70$.
Therefore any changes in $\epsilon$ that are consistent with a measurement
of $d_L (z=1)$ precise to 10 per cent do not change the sign of $q_0$,
at least for this fiducial model and form of perturbation. 

Such numbers,
although interesting, are of limited use since the form of 
the modification that we use 
is so specific.  We would, however, make the following two comments: 
\vskip 1pt
\noindent
(i) the choice of the fiducial model is important.  For fiducial models
like Model B which have $q_0$ close to zero, a small change in
$d_L (z=1)$ could be consistent with a model in which the sign of
$q_0$ changes.  
However, for model C, which is consistent with current observation, the
fiducial value of $q_0$ is far enough away from zero that all of 
the modifications considered here 
do not change the sign of $q_0$
without changing $d_L (z=1)$ at the
10\% level. 
\vskip 1pt
\noindent
(ii) small dependences of the Ricci scalar on any metric coordinates introduce
small changes in the Friedmann equation.  Larger dependances that are power
laws in either time or in the radial position coordinate also have
negligible effect.   Larger dependencies 
having other functional forms will have
a bigger effect but these will probably need to be fairly finely tuned if
they are going to affect the relationship between $d_L(z)$ and $q_0$ and
not have any appreciable signature in either the galaxy distribution 
on large scales or
the cosmic microwave background; 
\vskip 1pt
\noindent
(iii) invalidating general relativity in the way defined in Table 1
does not seem to have a strong
effect on the interplay between $d_L(z=1)$ and $q_0$.   Values of $\epsilon$
large enough to change the sign of $q_0$ would cause a very substantial
change in $d_L(z=1)$ (except for Model B, but see point (i)); 
\vskip 1pt
\noindent
(iv) 
Modifying the equation of state according to the quintessence
formulation has some 
effect on the interplay between $d_L(z=1)$ and $q_0$. 
Maor et al.~(2000)
consider the case of a time-varying equation of state, which is outside
the bounds of the simple mathematical treatment presented here, and find
that relaxing the assumption of a constant equation of state can have very
profound implications for using $d_L (z)$ to determine $q_0$ and more
generally the fate of the Universe.  

In these calculations, we have concentrated on measurements of $d_L(z)$
at $z=1$.  In principle, measurements at higher redshifts would help since  
$ {{\partial d_L (z)}\over{\partial q_0}}$ 
increases as $z$ increases (formally, this is
true only if the underlying cosmology is known
{\it a priori}; see also the discussion on this
point in Maor et al.~2000).  But one would then have additional systematic
effects that would need to be understood, like K-corrections of Type Ia
supernovae, complications due to dust extinction (Riess et al.~2000) and
possibly even gravitational lensing by matter along our line of
sight (Metcalf \& Silk 1999).  
More precise measurements at $z=1$ and lower 
redshifts would also help, although
these may be fundamentally limited by intrinsic scatter in the 
peak luminosity of Type Ia supernovae (see the references in the first
paragraph of Section 1). 

\section{Summary}

Therefore the relationship between $d_L(z=1)$ and
the sign of $q_0$ is fairly
robust given the analysis presented here unless the value of
$q_0$ is close to zero.  
For the $\Omega_{\Lambda} = 1 - \Omega_{\rm M} = 0.8$ model currently
favoured by observation, the value of $q_0$ is far enough away from
zero that this concern is not valid. 
This analysis is, however, limited, in that it only considers modifications
to the Friedmann models of a very specific type, and those individually. 
The joint effects of two or more modifications 
might be more dramatic, but at
this stage there are too many unconstrained ways to formulate these effects
for a detailed analysis to be productive.
In the long term, measurements
like those of the cosmic microwave background (Efstathiou et al.~1999), 
gravitational lensing
statistics (e.g.~Waga \& Frieman 2000) 
and galaxy number counts (Newman \& Davis 2000)
will provide other constraints.  

In order to address
formally
the question as to whether or not the Universe is accelerating, it will
be necessary to have measurements which probe the second time
derivative of $R(t)$.  However other, less direct,
measurements like the ones
listed in the previous
paragraph and the supernova measurements will provide powerful hints
if the association between the luminosity distance-redshift relation  
and the sign of $q_0$ really does turn out to be only
weakly model-dependant.

\end{document}